\documentstyle[11pt,epsf]{article}
\begin{document}
\catcode`@=11
\long\def\@caption#1[#2]#3{\par\addcontentsline{\csname
  ext@#1\endcsname}{#1}{\protect\numberline{\csname
  the#1\endcsname}{\ignorespaces #2}}\begingroup
    \small
    \@parboxrestore
    \@makecaption{\csname fnum@#1\endcsname}{\ignorespaces #3}\par
  \endgroup}
\catcode`@=12
\def\marginnote#1{}
\newcommand{\newc}{\newcommand}
\newc{\mtop}{\mt}
\newc{\btau}{$b$--$\tau$}
\newc{\bino}{\widetilde B}
\newc{\wino}{\widetilde W}
\newc{\beq}{\begin{equation}}
\newc{\eeq}{\end{equation}}
\newc{\bea}{\begin{eqnarray}}
\newc{\eea}{\end{eqnarray}}
\newc{\onehalf}{\frac{1}{2}}
\newc{\gsim}{\lower.7ex\hbox{$\;\stackrel{\textstyle>}{\sim}\;$}}
\newc{\lsim}{\lower.7ex\hbox{$\;\stackrel{\textstyle<}{\sim}\;$}}
\newc{\alphas}{\alpha_s}
\newc{\tanb}{\tan\beta}
\newc{\mz}{m_Z}		\newc{\mw}{m_W}
\newc{\mhalf}{m_{1/2}}
\newc{\mzero}{m_0}
\newc{\muzero}{\mu_0}
\newc{\sgnmu}{{\rm sgn}\,\muzero}
\newc{\azero}{A_0}
\newc{\Atop}{A_t}
\newc{\bzero}{B_0}
\newc{\mt}{m_t}
\newc{\mb}{m_b}
\newc{\mtau}{m_\tau}
\newc{\mq}{m_q}
\newc{\htop}{h_t}
\newc{\hbot}{h_b}
\newc{\htau}{h_\tau}
\newc{\mtpole}{M_t}
\newc{\mbpole}{M_b}
\newc{\mqpole}{M_q}
\newc{\mgut}{M_X}
\newc{\mx}{\mgut}
\newc{\alphax}{\alpha_X}
\newc{\ie}{{\it i.e.}}
\newc{\etal}{{\it et al.}}
\newc{\eg}{{\it e.g.}}
\newc{\etc}{{\it etc.}}
\newc{\hh}{{h^0}}
\newc{\mhh}{m_\hh}
\newc{\hH}{{H^0}}
\newc{\mhH}{m_\hH}
\newc{\hA}{{A^0}}
\newc{\mhA}{m_\hA}
\newc{\hpm}{{H^\pm}}
\newc{\mhpm}{m_\hpm}
\newc{\stp}{{\widetilde t}}
\newc{\stopl}{{\stp_L}}
\newc{\stopr}{{\stp_R}}
\newc{\stau}{{\widetilde\tau}}
\newc{\gluino}{{\widetilde g}}	\newc{\mgluino}{m_{\gluino}}
\newc{\MS}{{\rm\overline{MS}}}
\newc{\DR}{{\rm\overline{DR}}}
\newc{\ev}{{\rm\,eV}}
\newc{\gev}{{\rm\,GeV}}
\newc{\tev}{{\rm\,TeV}}
\newc{\bsg}{BR$(b\to s\gamma)$}
\newc{\abundchi}{\Omega_\chi h_0^2}
\newc{\mchi}{m_\chi}
\newc{\mcharone}{m_{\charone}}	\newc{\charone}{\chi_1^\pm}
\newc{\mneutone}{m_{\neutone}}	\newc{\neutone}{\chi^0_1}
\newc{\mneuttwo}{m_{\neuttwo}}	\newc{\neuttwo}{\chi^0_2}
\def\NPB#1#2#3{Nucl. Phys. {\bf B#1} (19#2) #3}
\def\PLB#1#2#3{Phys. Lett. {\bf B#1} (19#2) #3}
\def\PLBold#1#2#3{Phys. Lett. {\bf#1B} (19#2) #3}
\def\PRD#1#2#3{Phys. Rev. {\bf D#1} (19#2) #3}
\def\PRL#1#2#3{Phys. Rev. Lett. {\bf#1} (19#2) #3}
\def\PRT#1#2#3{Phys. Rep. {\bf#1} (19#2) #3}
\def\ARAA#1#2#3{Ann. Rev. Astron. Astrophys. {\bf#1} (19#2) #3}
\def\ARNP#1#2#3{Ann. Rev. Nucl. Part. Sci. {\bf#1} (19#2) #3}
\def\MODA#1#2#3{Mod. Phys. Lett. {\bf A#1} (19#2) #3}
\def\ZPC#1#2#3{Zeit. f\"ur Physik {\bf C#1} (19#2) #3}
\def\APJ#1#2#3{Ap. J. {\bf#1} (19#2) #3}
\begin{titlepage}
\begin{flushright}
{\large
UM-TH-94-03\\
February 1994\\
}
\end{flushright}
\vskip 2cm
\begin{center}
{\large\bf PREDICTIONS FOR CONSTRAINED MINIMAL SUPERSYMMETRY WITH
BOTTOM-TAU MASS UNIFICATION}
\vskip 1cm
{\large
Chris Kolda\footnote{E-mail: {\tt kolda@umich.edu}},
Leszek Roszkowski\footnote{E-mail: {\tt leszekr@umich.edu}},
James D. Wells\footnote{E-mail: {\tt jwells@umich.edu}}
and
G.L. Kane\footnote{E-mail: {\tt gkane@umich.edu}},
\\}
\vskip 2pt
{\it Randall Physics Laboratory, University of Michigan,\\ Ann Arbor,
MI 48190, USA}\\
\end{center}
\vskip .5cm
\begin{abstract}
We examine the Constrained Minimal Supersymmetric Standard
Model (CMSSM) with an additional requirement of strict
\btau\ unification in the region of small $\tanb$.
We find that the parameter space becomes completely limited
below about 1\tev\ by physical constraints alone,
without a fine-tuning constraint. We study the
resulting phenomenological consequences, and
point out several ways of falsifying the adopted \btau\ unification
assumption. We also comment on the effect of a constraint from the
non-observation of proton decay.
\end{abstract}
\end{titlepage}
\setcounter{footnote}{0}
\setcounter{page}{2}
\setcounter{section}{0}
\newpage

\section{Introduction}
A recent resurgence of strong interest in supersymmetry (SUSY) has
led to a number of attempts at exploring the physical spectra and
phenomenological consequences associated with the Minimal Supersymmetric
Standard Model (MSSM) in the context of Grand Unified Theories (GUTs).
This renewed interest is due primarily to the realization that gauge
coupling unification within the Standard Model
(SM) does not occur at any scale as one would
expect from GUTs such
as $SU(5)$. On the other hand, within the MSSM such unification
is found to be possible.
Early studies~\cite{early} have been followed by a series of
increasingly elaborate, and increasingly precise, analyses which have built
complete SUSY spectra consistent with the unification assumption (see
\cite{kkrw1} and references therein). These studies
have mostly used the well-motivated supergravity (SUGRA)
assumptions which suggest equating many of the unknown soft
SUSY-breaking mass terms at the GUT scale. Within
this framework, very complete studies can be done covering the entire range of
possible SUSY masses, and specific, testable predictions can be made.

In a previous work~\cite{kkrw1}, we have examined the MSSM
under a number of general assumptions about the unification of the gauge
couplings and masses, independent of the choice of gauge unification
group.\footnote{We only require that the choice of unification group lead
to $\sin^2\theta_{\rm w}(\mx)=\frac{3}{8}$ which
also holds in many phenomenologically viable superstring-derived models.}
Within this context, a number of
predictions, bounds, and signals were deduced and studied.
A specific choice of GUT model and/or further assumptions or constraints
could only serve to sharpen these predictions.

In this letter we consider one further aspect of unification: the apparent
unification of the bottom quark and tau lepton Yukawa couplings at the GUT
scale, as would be expected in a GUT containing minimal $SU(5)$ Yukawa
interactions. Semi-analytic studies completed recently by several
groups~\cite{kkrw1,bbop,cpw,lp2} have
found, however, that the MSSM does not in general produce this \btau\
mass unification except in small and well-defined regions of the parameter
space of $\mt$ and $\tanb$. Specifically, it
has been realized that, up to GUT-scale threshold corrections,
\btau\ mass unification can only occur if the top Yukawa coupling
is at or near its infrared pseudo-fixed point.

In Ref.~\cite{kkrw1}, we examined the size of the GUT-scale corrections
necessary in order to remove the strong constraints on $\tanb$ and $\mtop$.
We found that ${\cal O}(10\%)$
corrections both in the gauge and Yukawa unification were more than
adequate for allowing \btau\ mass unification over very wide ranges of
values for $\tanb$ and $\mtop$. Nonetheless, examinations of ``typical''
GUT threshold corrections~\cite{lp2} have yielded corrections too small
to significantly alter the relation between
$\tanb$ and $\mtop$, so a detailed exploration of the
SUSY parameter space indicated by \btau\ mass unification seems well-motivated.
Our goal is to derive testable experimental consequences of this assumption,
and to point out how \btau\ mass
unification within the MSSM can be falsified in a number of ways.
Finally, we will comment on the effects of a constraint from the
non-observation of the proton decay.

\section{The Pseudo-Fixed Point Solutions}

It was recognized several years ago that one of the strengths of the MSSM
over the SM was that in the MSSM the
$b$- and $\tau$-Yukawa running couplings meet at roughly the same mass scale at
which the unification of the gauge couplings takes place~\cite{bothunif}.
As both the experimental data and the theoretical
calculations became more precise, it became clear that gauge coupling
unification within the MSSM occurs over the entire range of theoretically
acceptable SUSY mass scales. At the same
time, however, Yukawa unification within the MSSM is not so trivial. For most
choices of input parameters (\eg, $\tanb$ and $\mtop$), the $\tau$-Yukawa
coupling is as much as ${\cal O}(20\%)$ larger than that of the
$b$-quark at the gauge coupling unification scale.
And because the slopes of the bottom and tau Yukawas are typically
flat at large scales, the
Yukawa couplings can ``unify'' (\ie, cross) at a mass scale many orders of
magnitude smaller than the GUT scale, even though their GUT-scale values
may differ by only 10--20\%.

The exception to this general trend, however, occurs when either
{\em (i)} the top Yukawa coupling, or {\em (ii)} the $b$- and $\tau$-Yukawa
couplings, are at or very near their infrared pseudo-fixed point values.
That pseudo-fixed point value is the value of the Yukawa coupling (at the
electroweak scale) which produces a Landau pole precisely at the GUT scale;
that is, values of the coupling greater than the fixed point value will
become non-perturbative at scales below $\mx$
when run up using its renormalization group equations (RGEs).

The values for the top, bottom, and tau Yukawa couplings corresponding to the
pseudo-fixed point can be precisely determined.
For the case at hand, one finds~\cite{bbop,cpw}
that there are two conditions for
sitting on or near one of the pseudo-fixed points (that is, for finding
\btau\ Yukawa unification) either one of which must be satisfied. Either
\beq
\mtpole\simeq (200\gev)\sin\beta
\label{mtop:eq}
\eeq
\noindent or
\beq
50\lsim\tanb\lsim70.
\label{tanb:eq}
\eeq
The first condition corresponds to the top Yukawa pseudo-fixed point, while the
second is the bottom-tau Yukawa pseudo-fixed point. Also,
because the difference
between the top quark pole mass and running ($\MS$ or $\DR$)
mass can be as much as $10\gev$ in the region of interest, one must be careful
to specify to which top quark mass one is referring. We will use
$\mq$ to specify a quark running mass (or a general definition
when the choice is irrelevant) and $\mqpole$ its pole mass.

\begin{figure}
\centering
\epsfysize=3.25in
\hspace*{0in}
\epsffile{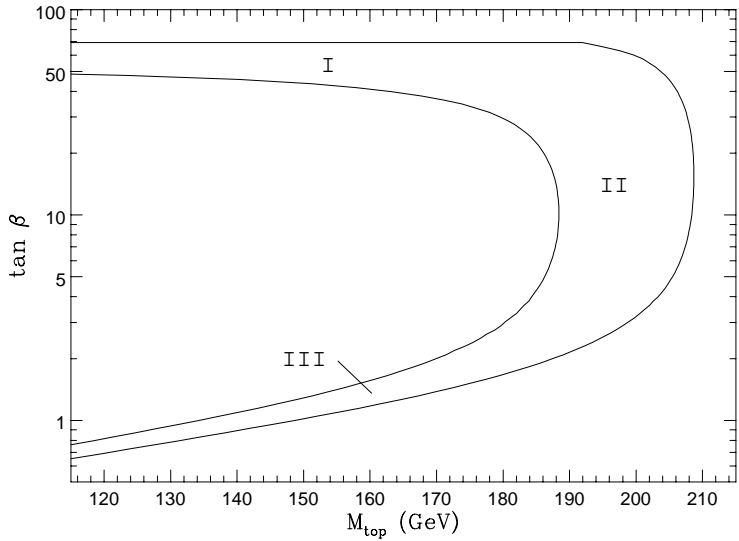}
\caption{The $(\mtpole,\tanb)$ plane showing the region consistent with \btau\
unification. The width of the region is due to the $3\sigma$ uncertainty in
$\mbpole$ and the $1\sigma$ uncertainty in $\alphas$ (see text).
The three regions
of unification are all visible: (I) $\tanb\simeq50\sim70$,
(II) $190\gev\lsim\mtpole\lsim210\gev$, and (III) $\tanb\simeq1$.
}
\label{tanb:fig}
\end{figure}

In Fig.~\ref{tanb:fig}
we have shown the region in the $(\mtpole,\tanb)$ plane consistent
with \btau\ mass unification. The width of the region is due to the
$3\sigma$ uncertainty in the $b$-quark mass (using $\mbpole=4.9\pm0.07\gev$
\cite{mbot})
and the $1\sigma$ uncertainty in $\alphas(\mz)$
($\alphas(\mz)=0.120\pm0.007$~\cite{lepreview});
in this figure none of the width comes from GUT threshold corrections to
the Yukawa unification.

{}From Fig.~\ref{tanb:fig}
it is evident that there are three primary regions of interest in this plane.
The first is the region (labelled {\small I}) of large $\tanb$, over all
$\mtpole$, where the $b$- and $\tau$-Yukawa couplings reach their pseudo-fixed
points. Some part of this region, corresponding also to large $\mtpole$, is of
particular interest to those studying $SO(10)$ unification with a minimal
Yukawa sector. There one can obtain the GUT-scale prediction $\mb=\mtau=
\mtop$, which when renormalized at the electroweak scale yields
$\tanb\simeq\frac{\mtop}{\mb}\simeq50\sim70$. Because of that relation,
this region deserves consideration and some studies have been carried
out~\cite{largetb,hrs,copw:so10}.
However, certain difficulties invariably arise in considering the MSSM
with very large $\tanb$. Three in particular stand out.

First, it has been argued that within the MSSM (with two Higgs doublets)
large $\tanb$ is unnatural~\cite{nelson}. Specifically, one finds that
large hierarchies,
only some of which can be protected by additional imposed symmetries, arise
among the mass parameters of the Higgs potential.
Second, one has particular difficulty with the 1-loop corrections to the
$b$-quark mass, which are proportional to $\tanb$ and
can easily be of ${\cal O}(50\%)$ unless new symmetries are imposed in order
to control them~\cite{hrs}.
This issue has been more recently investigated in Ref.~\cite{copw:so10}
in the context of an $SO(10)$ GUT model and it has been argued that requiring
dynamical electroweak symmetry breaking makes those corrections well defined
though generally unsuppressed.
Finally, models with large $\tanb$ tend to produce very small branching
ratios for the FCNC process $b\to s\gamma$ due to their suppressed charged
Higgs contributions, perhaps inconsistent with recent
CLEO data. Therefore, we choose to put off any further consideration
of this region for now.

The second region (labelled {\small II} in Fig.~\ref{tanb:fig})
leading to \btau\ mass unification is found at large $\mtpole$ between
about $190$ and $210\gev$ for almost all values
of $\tanb$ (between about 2 and 60).
This region of parameter space is strongly disfavored by the $2\sigma$ LEP
limit $\mtop\lsim 180\gev$ for $\mhh<120\gev$~\cite{lp3}.
Further, if FNAL does
see top quark events, one could exclude top quark masses in this region
due to the small cross section expected for such large $\mt$.
Therefore, we will not consider this region in the present study.

The third and final region (labelled {\small III} in Fig.~\ref{tanb:fig})
is at low $\tanb$, with $\mtpole\lsim
190\gev$, where there is almost a one-to-one correspondence between
$\tanb$ and $\mtpole$. This range is perhaps of more immediate interest
since it corresponds to the top mass range to be covered by the Tevatron.
And, having disfavored the previous two regions, we
are led to consider in detail this region of $\tanb$ near one, for
$155\gev\leq\mtpole\leq185\gev$.
As we will show below, this scenario leads to very specific and falsifiable
predictions which could rule it out.

One should remark on the size of the radiative corrections to the $b$-quark
mass in the low $\tanb$ regime.
We find for all solutions in this third region that the
(leading) gluino-induced corrections to the bottom quark mass are always
small, $\delta m_b/m_b\lsim2\%$; the
higgsino-induced corrections are approximately four times smaller still. These
corrections are far too small to alter our results. We also find that the
sizes of these corrections depend only very weakly on the SUSY mass parameters
$\mzero$, $\mhalf$, and $\mu$ in fully consistent solutions (see following
section), and so remain small even at large SUSY mass scales.

\section[]{The Constrained Minimal Supersymmetric \\ Standard Model (CMSSM)}

In Ref.~\cite{kkrw1}, we described the construction of what we termed the
Constrained MSSM (or CMSSM),
built by relating the MSSM soft-breaking terms
through the minimal SUGRA assumptions and then constraining these solutions
as summarized below. Here we briefly summarize the basic points of that
construction.

For each choice of $\mtpole$, $\mzero$, $\mhalf$, $\azero$, and $\sgnmu$
($\tanb$ is now determined through the requirement of \btau\ unification
as described earlier) we find a solution in the CMSSM; each solution is a
complete set of values for $\alphas$ and the masses of the Higgs bosons and
all the superpartners. (The exact procedure for building such complete
solutions is summarized in Ref.~\cite{kkrw1}.) If the Higgs potential at
the electroweak scale does not admit electroweak symmetry breaking (EWSB)
consistent with the SM, that choice of input parameters is discarded.
Further constraints are then applied.

Besides requiring that EWSB occur, we demand that all physical
mass-squares remain positive. We demand that all
solutions be unobservable by current direct experimental searches, including
the requirement that the solutions provide a \bsg\ consistent with CLEO
data. Furthermore, we calculate
the relic density of the lightest SUSY particle (LSP), demanding only that the
LSP be neutral, and, from limits on the age of the universe of 10 billion
years, we demand that $\Omega_{LSP}h_0^2<1$.
Those solutions which finally remain after
all these cuts comprise the allowed parameter space of the CMSSM.

\section{Results}

We have examined solutions for three representative top quark (pole) masses of
$\mtpole=$155, 170, and 185$\gev$ in order to study the range that seems to be
indicated by the LEP analysis and early CDF results. In building the parameter
space of solutions, we have varied $\mzero$ and $\mhalf$ from $20\gev$
to the TeV
scale logarithmically; no upper bound is set by hand on $\mzero$ or
$\mhalf$, their upper bounds coming from the constraints on the solutions
which define the CMSSM. The lower bound of $\mzero$ was not taken
to zero, but values in the lower region
($\mzero\simeq{\cal O}(20\gev)$) are for all purposes phenomenologically
identical to each other as well as to No-Scale models, since
experimental bounds force $\mhalf\gsim80\gev$.
The value of $\azero$ was restricted to the range $-2.5\leq\azero/\mzero\leq
2.5$ in order to avoid possible color-breaking minima of the
full scalar potential. The final free parameter, $\sgnmu$, was allowed
to take both possible values.

\subsection{Allowed Parameter Space and Mass Spectra}

One of the remarkable consequences of considering this low $\tanb$ limit
with a complete analysis is the existence of upper
bounds on the mass parameters of the MSSM {\em without resorting to an
imposed arbitrary fine-tuning condition}. In particular the combination of
various mass bounds from direct searches, the age of the universe constraint,
and the requirement that the LSP be electrically neutral combine to put strict
upper (and lower) bounds on $\mzero$ and $\mhalf$, and therefore on the
mass spectrum of the MSSM. These bounds are entirely due to the physical
constraints of the CMSSM.
As we pointed out previously~\cite{kkrw1}, for small $\tanb\lsim2$ and/or
large $\mtop\gsim 170\gev$,
the parameters $\mhalf$ and $\mzero$ (and therefore the whole SUSY spectrum)
are completely constrained from above in the ${\cal O}(1\tev)$ range.
The case considered here falls into that category.
These absolute upper bounds are $\mtop$-dependent and are
usually somewhat weaker than those imposed by simply requiring all SUSY
masses below about 1~\tev, or than those which the fine-tuning constraint we
used in Ref.~\cite{kkrw1} would have
permitted. For this reason, in examining some phenomenological applications
of the solutions in the CMSSM, we will place an additional
fine-tuning constraint. We do so in order to ensure
that phenomenological results of this study are ``realistic''; that is,
although consistent solutions may exist with large SUSY mass scales,
we wish to exclude these from phenomenological consideration on the basis
that they reintroduce too much fine-tuning into the physics.
(Our precise definition of fine-tuning is discussed in Ref.~\cite{kkrw1}.
We note that the definition that we use diverges at
$\tanb=1$; the values of $\tanb$ that we consider here are far enough
from unity so that this effect is not significant, and becomes irrelevant
as $\tanb$ increases with increasing $\mtop$.)

\begin{figure}
\centering
\epsfxsize=5.75in
\hspace*{0in}
\epsffile{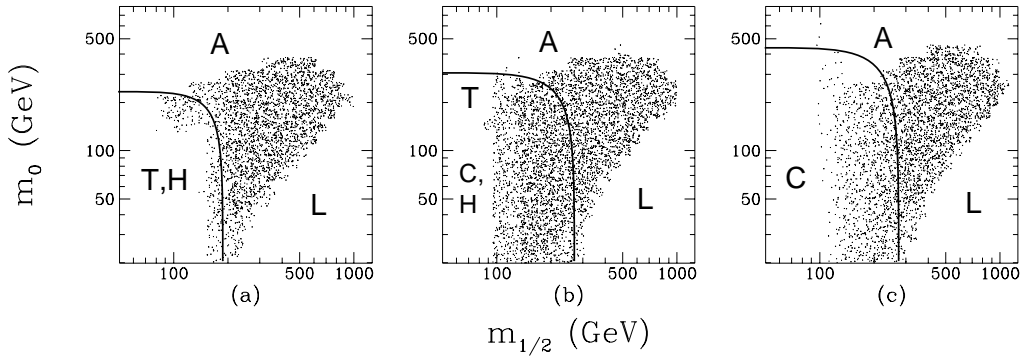}
\caption{The regions of the $(\mhalf,\mzero)$ plane consistent with
low $\tanb$ \btau\ mass unification, given all the constraints of the
CMSSM. For (a) $\mtpole=155\gev$, (b) $\mtpole=170\gev$ and (c)
$\mtpole=185\gev$.
Each dot represents one solution belonging to the CMSSM. Because
$\mzero$ and $\mhalf$ are discrete inputs in this approach, the points have
been ``smeared'' to show variations in density. The regions with no
solutions are labelled as to which constraints forbid solutions there:
age of universe bound ({\bf A}), neutral LSP requirement ({\bf L}),
non-tachyonic stop ({\bf T}); Higgs mass bound ({\bf H}), and chargino
mass bound ({\bf C}). The solid curves
in each figure are the approximate cutoffs dictated by our choice of
fine-tuning constraint.
}
\label{envelopes:fig}
\end{figure}

In Fig.~\ref{envelopes:fig}
we have shown the allowed range of parameter space in the
$(\mhalf,\mzero)$ plane for $\mtpole=155$ (Fig.~\ref{envelopes:fig}a),
$170$ (Fig.~\ref{envelopes:fig}b),
and $185\gev$ (Fig.~\ref{envelopes:fig}c) without fine-tuning
constraints. All three graphs exhibit many similarities
which are general features of the CMSSM for all $\tanb$~\cite{kkrw1}.
In both cases the region of large $\mhalf\gg\mzero$ is excluded
by demanding a neutral LSP.
This uniquely selects the lightest neutralino $\chi\equiv\neutone$ (mostly a
bino-like state) for which we calculate the relic abundance. Then we find that
large $\mzero$ are excluded by requiring $\abundchi<1$.

For the $\mtpole=155\gev$ case, we find that Yukawa unification allows $\tanb$
to take on values only in the range
$1.1\lsim\tanb\lsim 1.4$. This, coupled with
the requirements of the CMSSM, restricts $80\gev\lsim\mhalf\lsim940\gev$ and
$\mzero\lsim350\gev$.
When we apply the fine-tuning constraint $f\leq50$, we
find the approximate bound shown as a solid line in
Fig.~\ref{envelopes:fig}a. For this
subset of solutions, we find new upper bounds on the parameters of the model.
In particular, the fine-tuning constraint yields $\tanb\lsim1.2$, $\mhalf\lsim
180\gev$ and $\mzero\lsim210\gev$. Notice also the strong lower bounds placed
on the parameter space. This bound comes from two sources:
for $\mu>0$, solutions with small $\mhalf$,$\mzero$ are
ruled out by the Higgs mass bound; for $\mu<0$, it is the $\stp_1$ mass bound
that rules out the same approximate region. This effect is strongly diminished
with increasing $\tanb$ (and therefore $\mtpole$).

The $\mtpole=185\gev$ case in Fig.~\ref{envelopes:fig}c
exhibits one additional interesting feature.
For a relatively narrow range of $\mhalf\simeq100\gev$ solutions with much
larger $\mzero$ are allowed. The source of this behavior is the $Z$-pole
enhanced neutralino pair annihilation
into a pair of ordinary fermions. (For $\mhalf\simeq100\gev$, $\mchi\simeq
0.45\, \mhalf\simeq\mz/2$.) This exchange vanishes in the limit $\tanb=1$ and
is therefore negligible in Fig.~\ref{envelopes:fig}a where $\tanb\lsim1.4$.
For $\mtpole=185\gev$, we find $1.8\lsim\tanb\lsim 3.1$ and the effect
of the $Z$-exchange becomes important.
Our analysis provides an overall envelope of
$100\gev\lsim\mhalf\lsim1.1\tev$ and $\mzero\lsim600\gev$.
Once again the fine-tuning constraint tends to lower the upper bounds of
the various model parameters. In particular, we find $\tanb\lsim2.4$,
$\mhalf\lsim290\gev$ and $\mzero\lsim420\gev$ after applying the fine-tuning
constraint. The corresponding fine-tuning bound is shown as a solid
line in Fig.~\ref{envelopes:fig}c.

The intermediate case for $\mtpole=170\gev$ in Fig.~\ref{envelopes:fig}b does
demonstrate some hint of the $Z$-pole effect found at the larger $\tanb$
associated with $\mtpole=185\gev$. Here we find $1.4\lsim\tanb\lsim 1.9$,
$90\gev\lsim\mhalf\lsim940\gev$, and $\mzero\lsim500\gev$ without a
fine-tuning constraint. Upper
bounds with the fine-tuning constraint are modified to be $\tanb\lsim1.5$,
$\mhalf\lsim260\gev$ and $\mzero\lsim300\gev$.

It is worth noting that the region $\mzero\gg\mhalf\simeq\mz$
(and small $\tanb$) is favored by the non-observation of proton
decay in $SU(5)$-based GUTs with minimal Higgs sector~\cite{an}.
In this case the neutralino relic abundance is reduced by the $Z$-
and $\hh$-pole effects. In this region the LSP still remains mostly
bino-like, although with somewhat smaller bino component ($\gsim90\%$).
However, predictions for proton decay can be suppressed with more
complicated Higgs sectors (see, \eg, Ref.~\cite{babubar}) and, since
we do not assume any specific GUT model here, we will also not use this
constraint to limit the parameter space of the CMSSM. We also note that
our numerical routine for the relic abundance is not designed to properly
calculate $\abundchi$ in the vicinity of a pole (that is, within
about 10\gev\ of the pole) and therefore the values of $\abundchi$ in this
regions are only indicative.

One should note for all bounds throughout this study that the
exact values depend on our numerical sampling of the original input parameter
space and so should be considered with appropriate errors. In particular,
upper (lower) bounds on
$\mhalf$ could be modified upwards (downwards) by as much as 12\% with
a smaller sampling grid; bounds on $\mzero$ could likewise be increased
(decreased) by as much as 20\%.

Because such strict bounds exist for these cases (with and without
fine-tuning), we can place bounds on the CMSSM mass spectra.
In Table~\ref{bounds:table} the numerical bounds on a variety of
important quantities are shown for all
three top masses, with and without the fine-tuning constraint.

\begin{table}
\centering
{\small
\begin{tabular}{|c||c|c|c||c|c|c||c|c|c|}
\hline
Mass Limits & \multicolumn{3}{c||} {$\mtpole=155\gev$}  &
\multicolumn{3}{c||}{$170\gev$} &
\multicolumn{3}{c|} {$185\gev$} \\ \cline{2-10}
(GeV) & lower & upper & FT & lower & upper & FT & lower & upper
& FT  \\
\hline\hline
$\mhalf$ & 80 & 940 & 180 & 90 & 940 & 260 & 100 & 1060 & 290 \\ \hline
$\mzero$ & 0 & 350 & 210 & 0 & 500 & 300 & 0 & 600 & 420 \\ \hline
$|\mu(\mz)|$ & 520 & 1800 & 660 & 250 & 1470 & 560 & 210 & 1390 & 520 \\ \hline
$M_2$ & 65 & 780 & 150 & 70 & 780 & 210 & 80 & 880 & 240 \\ \hline
$\hh$ & 60 & 105 & 81 & 60 & 124 & 99 & 75 & 149 & 118 \\ \hline
$\hA$ & 730 & 2550 & 925 & 330 & 2040 & 770 & 260 & 1910 & 710 \\ \hline
$\stp_1$ & 38 & 1340 & 215 & 38 & 1400 & 470 & 115 & 1550 & 510 \\ \hline
$\stau_1$ & 65 & 420 & 210 & 50 & 500 & 300 & 55 & 600 & 420 \\ \hline
$\widetilde{l_L}$ & 115 & 670 & 215 & 65 & 710 & 310 & 70 & 780 & 430 \\ \hline
$\widetilde q$ & 235 & 1740 & 425 & 245 & 1870 & 590 & 255 & 2100 & 670 \\
\hline
$\gluino$ & 215 & 2030 & 450 & 240 & 2040 & 630 & 270 & 2300 & 710 \\ \hline
$\chi=$\/LSP  & 35 & 410 & 70 & 25 & 410 & 110 & 25 & 465 & 125 \\ \hline
$\chi_1^\pm\simeq\chi_2^0$ & 75 & 770 & 135 & 48 & 780 & 220 & 48 & 875 & 245
\\ \hline\hline
$\tanb$ & 1.1 & 1.4 & 1.2 & 1.4 & 1.9 & 1.5 & 1.8 & 3.1 & 2.4 \\ \hline
$\alphas(\mz)$ & 0.117 & 0.125 & $0.121^\dagger$ & 0.122 & 0.129
& $0.124^\dagger$ & 0.124 & 0.133 & $0.124^\dagger$ \\ \hline
\multicolumn{10}{l}{\footnotesize ${}^\dagger$~Lower bound with FT constraint.}
\\
\end{tabular}}
\caption{The bounds of the masses and parameters of the
MSSM under the constraint of low $\tanb$ Yukawa unification for
$\mtpole=155$, $170$, and $185\gev$. For each top quark mass the lower
bound (labelled {\em lower}), the absolute upper bound with no
fine-tuning constraint ({\em upper}) and the upper bound with the
fine-tuning constraint ({\em FT\/}) are shown.
All masses are in GeV. The general
masses $\widetilde q$ and $\widetilde{l_L}$ represent the bounds on all
squarks and LH-sleptons excluding the third generation. The bounds on all
first and second generation RH-sleptons are essentially those of the
$\stau_1$ for low $\tanb$. Note that there is sensitivity to the
grid of values for $(\mzero,\mhalf,\azero)$ that we have chosen, and
therefore some uncertainty in the exact bounds.
}
\label{bounds:table}
\end{table}

It is significant that much of the region of low $\mhalf$ for the
$\mtpole=155\gev$ case has been excluded on
the basis of the non-discovery of the Higgs boson in direct searches. For all
solutions in the superunified MSSM, the lightest Higgs scalar ($\hh$)
has essentially identical properties to those of its SM counterpart;
that is $\sin^2(\beta-\alpha)
\approx1$ always. Therefore mass bounds on the SM Higgs apply equally well
to the $\hh$. (We conservatively require $\mhh>60\gev$.) However, $\hh$
receives large one-loop
radiative corrections which can increase its mass by ${\cal O}(\mz)$.
Therefore, limits on the solution space based on Higgs non-observation must
include the full one-loop radiative corrections, which in turn requires a full
calculation of the SUSY mass spectra, particularly the mass eigenstates of
the top squarks.

\begin{figure}
\centering
\epsfysize=2.75in
\hspace*{0in}
\epsffile{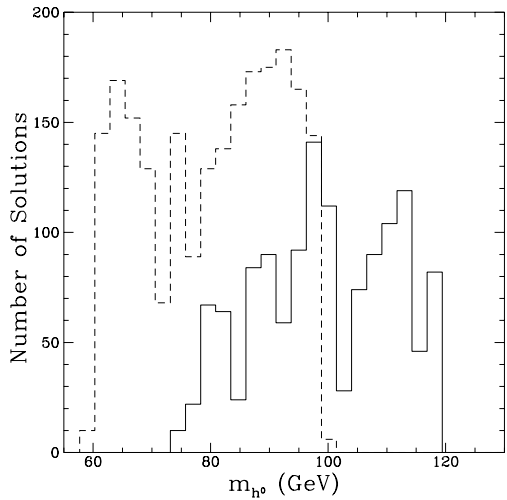}
\caption{The number of solutions with a given $\mhh$ versus the value of
$\mhh$, for $\mtpole=170\gev$ (dashed) and $185\gev$ (solid). Only the
solutions with small fine-tuning are plotted.
}
\label{higgs:fig}
\end{figure}

Because the top mass dependence of these radiative corrections is so strong,
one can clearly understand why the lower bound on $\mhalf$
due to the Higgs mass decreases with
increasing $\mt$; for larger $\mt$, one gets larger $\mhh$ from radiative
corrections, so the experimental bound on $\mhh$ rules out less of
parameter space. In Fig.~\ref{higgs:fig} we demonstrate
this behavior by plotting the number
of solutions leading to a given $\mhh$ versus $\mhh$, both for
$\mtpole=170\gev$ (dashed) and $185\gev$ (solid), subject to our fine-tuning
constraint.

There are also upper bounds on $\mhh$, and several groups have recently
examined these bounds within this approach~\cite{lp3,abs,flast}. However,
because in the $\tanb\to1$ limit $\mhh^2\to0$ at tree-level, these bounds
are highly dependent on the size of the radiative corrections to $\mhh^2$.
And because these corrections increase quartically with $\mtop$ and
logarithmically with the SUSY masses, their size is highly dependent on
one's assumptions about how to cut off the allowed MSSM parameter space,
that is, how one defines what ``too much fine-tuning'' means.

For $\mtpole=155$,
$170$, and $185\gev$, we find $\mhh$ to be less than $81$, $99$, and
$118\gev$ respectively, with the fine-tuning condition $f<50$ imposed. Without
this condition, the upper bounds increase, as the effective SUSY mass scale
is somewhat increased (compare Table~\ref{bounds:table}).
Though we believe that it is the first set of bounds that
should be taken as more indicative of our expectations
since they more fully contain theoretical prejudices which apply to SUSY,
one should bear in mind the strong dependence of these bounds on the
choice of fine-tuning condition. This caveat, however, does not apply to
the maximum values as given in Table~\ref{bounds:table}, where no
fine-tuning condition at all was used to bound the parameter space.
One should also note that calculations of the two-loop corrections to
$\mhh$ show a net decrease of $\mhh$ below its one-loop
value~\cite{hempfling}, and therefore will not disrupt our bounds.

Detection of the other Higgs bosons ($\hH$, $\hA$, and $\hpm$) cannot be
accomplished at LEP~II nor at the proposed NLC\/500~\cite{nlc500higgsrep}.
We find for $155\gev\leq\mtpole\leq185\gev$ that
$\mhH\simeq\mhA\simeq\mhpm\geq 260\gev$, outside the range of either
machine. We believe that the detection of the heavier Higgs scalar $\hH$
might be possible at the LHC somewhat beyond the asserted region
$\mhH\lsim2\mtpole$~\cite{higgsdetection}.
However, all solutions (with small fine-tuning) do have at least one
SUSY particle that is detectable at the LHC in addition to the $\hh$,
even if the heavy Higgs bosons are not.

Can LEP~II find any sparticles? We find that under the assumption
of low $\tanb$ \btau\ unification, detection of states other
than the lightest Higgs may be possible, but for a few solutions only.
Searches for light SUSY particles should concentrate on the
lighter stop ($\stp_1$), the lighter stau ($\stau_1$), the lighter chargino
($\chi_1^\pm$) and the second lightest neutralino ($\chi_2^0$, where
$m_{\chi_2^0}\simeq m_{\chi_1^\pm}$). Within the range of $\mtpole$
considered, we find solutions with masses for these particles all the
way down to their current experimental limits. In particular,
for such light stops there is a large mixing of the right-handed and
left-handed interaction eigenstates, so the simple approximation that
$\stp_1\simeq
\stp_R$ does not hold. Further studies of the detectability of the MSSM under
similar constraints and assumptions have been done in
Ref.~\cite{flast,gunion}.

We finally note that the resulting ranges of $\mgluino$ and scalar quark
masses are largely above the reach of the Tevatron. Thus finding any of those
particles well below the ranges indicated in Table~\ref{bounds:table} would
rule out the \btau\ unification if $\tanb$ is close to one.

\subsection{\bsg}

In the general superunified MSSM where Yukawa unification has
not been required, the recent CLEO bounds on \bsg\ have the ability to rule
out certain regions of parameter space and indicate a future ability to
further constrain or discover SUSY through more precise measurements of
\bsg~\cite{kkrw1}. However, one finds in
the region of low $\tanb$ Yukawa unification that almost all solutions
consistent with all other requirements of the CMSSM
naturally fall within the bounds of the CLEO data, and in particular,
no solutions provide larger branching ratios than are allowed by the data.
(We follow Ref.~\cite{bsg} for our calculations of the branching ratio.)
But as the following analysis emphasizes, this is not in general due simply to
the decoupling of the SUSY contributions; in fact, for solutions with low
fine-tuning, the SUSY and SM contributions are often comparable in size.

\begin{figure}
\centering
\epsfxsize=5.75in
\hspace*{0in}
\epsffile{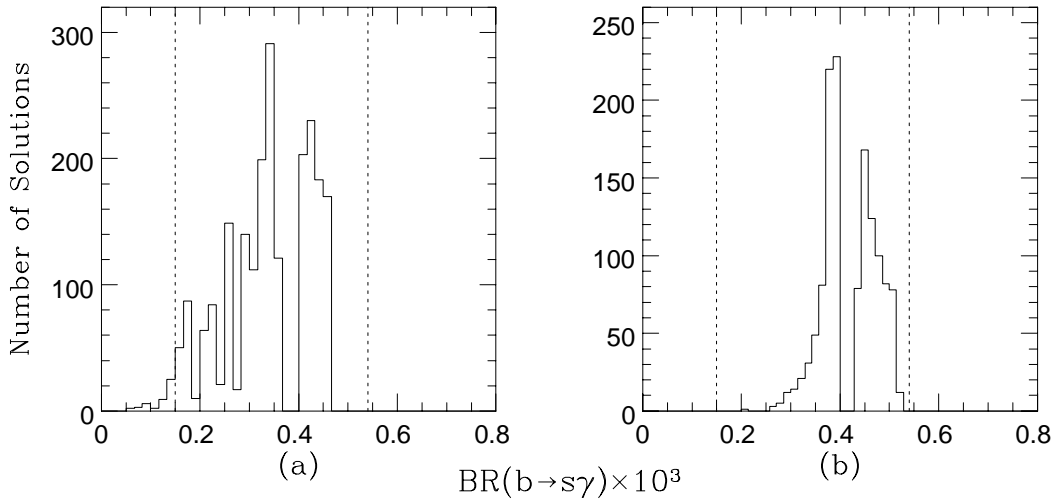}
\caption{Histogram of \bsg\ for (a)~$\mtpole=170\gev$ and
(b)~$\mtpole=185\gev$, for solutions corresponding to low fine-tuning.
}
\label{bsg:fig}
\end{figure}

Nonetheless, what is particularly noteworthy in the low $\tanb$ and small
fine-tuning limit is
that the branching ratio is highly dependent on the sign of $\mu$. In
Fig.~\ref{bsg:fig} we have histogrammed our calculated \bsg\ for
$\mtpole=170$ and $185\gev$ solutions with small fine-tuning (we comment
on relaxing the fine-tuning condition below).
The central peaked region falls approximately
at the SM prediction of the branching ratio, with larger $\mtpole$ moving
the peak (and the SM prediction) to larger values. Note however that the
histogram yields two separate, non-overlapping regions.
In both cases, the region of higher branching ratio is
occupied only with solutions of $\mu>0$. Likewise the region of lower
branching ratio is occupied only with solutions of $\mu<0$. Presuming that
both theoretical and experimental uncertainties to $b\to s\gamma$ ever become
small enough, one may then have a method by which to differentiate the sign of
$\mu$ through this process.

Why the two separate regions? Of the non-SM contributions to the
$b\to s\gamma$ amplitude, the dominant contribution
in this region of parameter space tends to come from the $\chi_1^\pm$--$\stp_1$
loop, with sign opposite to that of the SM $W^\pm$--$t$ contribution.
For all acceptable solutions of the CMSSM, one gets
a lighter chargino that is almost pure $\wino^\pm$ and so only has couplings to
$\stp_L$. Without mixing of the $\stp$'s
it is the $\stp_R$ that invariably comes out to be the lightest squark,
which leads to a very small $\chi_1^\pm$--$\stp_1$ contribution

Now it is well known that the $\stopl$ and $\stopr$ can have
large mixings proportional to $\mtop^2(\Atop+\mu/\tanb)^2$.
For small mixings the lighter stop
($\stp_1$) is almost pure $\stopr$; as the mixing increases $\stp_1$ gains a
larger $\stopl$ component and so the $\chi_1^\pm$--$\stp_1$ contribution
becomes sizeable. For solutions in which $\mzero$ is not too much
larger than $\mhalf$, one finds that $\Atop$ is driven negative at the
electroweak scale regardless of its magnitude and sign at the unification
scale. Therefore, if $\mu$ is also negative, large $\stopl$--$\stopr$
mixings result and $\stp_1$ has a sizeable $\stopl$ component; if
$\mu>0$, the $\Atop$ and $\mu$ contributions to the mixing
tend to cancel, forcing the $\stopl$ component of $\stp_1$ to be very small.

In the $\mu>0$ case, then, the $\chi_1^\pm$--$\stp_1$ coupling will
be very small, allowing the SM contribution to easily dominate. In the
$\mu<0$ case, however, the coupling will be sizeable, cancelling much or all
of the SM contribution. Therefore, the $\mu<0$ case will give much smaller
branching ratios than one would get from the
$\mu>0$ case, with the two regions separated near the SM prediction.

What happens as we allow larger $\tanb$ or larger fine-tuning?
Since the $\stopl$--$\stopr$ mixing goes as $(\Atop+\mu/\tanb)^2$,
it is clear that as $\tanb$ increases,
$\Atop$ will come to dominate the mixing and the results will be relatively
independent of $\sgnmu$. Similarly, the dependence on $\sgnmu$ disappears for
solutions corresponding to larger fine-tuning
because in this case $\mu$ typically becomes large
and its contribution dominates over that of $\Atop$, producing mixings
proportional only to $(\mtop\mu)^2$.
However, as expected, for solutions corresponding to large fine-tuning,
and therefore large masses in the loops,
the supersymmetric contributions go to zero, leaving only the SM
contribution and thus no $\sgnmu$ dependence anyway.

\subsection{The $\mu$-parameter}

Of particular interest is the Higgs/higgsino mass parameter $\mu$
which does not break SUSY and therefore could in principle take
values much larger than the soft SUSY-breaking parameters. In this approach,
however, its size is determined through the condition of electroweak symmetry
breaking and comes out to be of the same order of magnitude as $\mhalf$ and
$\mzero$, as has been discussed in detail in Ref.~\cite{kkrw1} and many
other places. Further potentially strong correlations can be derived by
imposing additional constraints or assumptions, like the \btau\ unification
discussed here.

Working in the top Yukawa pseudo-fixed point limit, the authors of
Ref.~\cite{copw1}
found semi-analytic expressions indicating a strong correlation
between $\mu$ and $\mhalf$ in the region of Yukawa
unification. Though their results were only at tree level and so did not
contain contributions from the one-loop effective potential, they are easily
extended to include the leading correction from the $\stp$ sector.
We find that for $\mhalf>\mzero,\mz$:
\beq
\mu^2\simeq\mhalf^2\,\frac{0.5+3.5\tan^2\beta}{\tan^2\beta-1} -
\frac{15\alpha_2}{2\pi}\left(\frac{\mtop^2}{\mw^2}\right)
\left[\log\left(\frac{5\mhalf^2}{\mz^2}\right)-1\right].
\label{ourfit:eq}
\eeq
The first term on the right in Eq.~(\ref{ourfit:eq}) is the tree level
contribution only~\cite{copw1}, while the second term represents the one-loop
leading correction.

\begin{figure}
\centering
\epsfxsize=5.75in
\hspace*{0in}
\epsffile{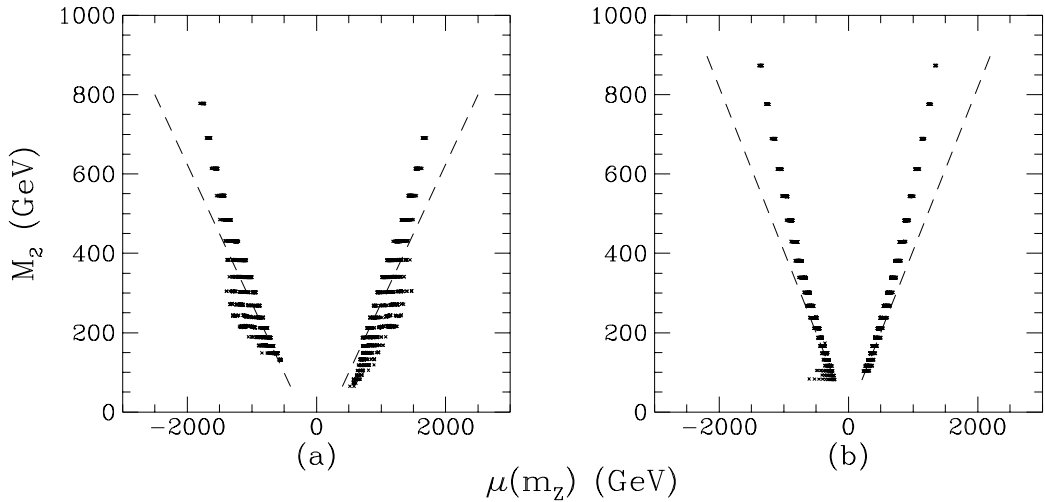}
\caption{Scatter plot of $M_2$ versus $\mu(\mz)$ for (a) $\mtpole=155\gev$ and
(b) $\mtpole=185\gev$ for all solutions in the CMSSM without a fine-tuning
constraint. Notice the ``$\mzero\gg\mhalf$'' points present in (b) allowed
because those solutions have LSP's falling in the pole of the $Z$-channel
annihilations. The tree level calculation of the $\mu$-$M_2$ correlation
is also shown (dashed line).
}
\label{mu:fig}
\end{figure}

We illustrate the behavior of $\mu$ as a function of $M_2$
($M_2\simeq0.8\,\mhalf$) in Fig.~\ref{mu:fig} for our solutions with
(a) $\mtpole=155\gev$ and (b) $\mtpole=185\gev$.
One can compare this to Fig.~31 of Ref.~\cite{kkrw1} where the
$(\mu,M_2)$ plane for general solutions in the CMSSM was displayed. There one
does not see evidence for the strong correlation between $\mu$ and $M_2$ (or
$\mhalf$) that one finds in the pseudo-fixed point limit. We have also
plotted in Fig.~\ref{mu:fig} a dashed line corresponding to the tree
level calculation of $\mu$ as given by the first term in
Eq.~(\ref{ourfit:eq}).

{}From Fig.~\ref{mu:fig} one sees that the tree level expression for
the $\mu$--$\mhalf$ correlation describes our solutions well until $M_2\gsim
400\gev$, where the slope rises due to the one-loop corrections.
The one-loop effects are large enough so that the tree level calculation
for $\mu$ at $\mhalf\simeq1\tev$ is ${\cal O}(50\%)$ larger than
the actual one-loop value.

One should note, however, that there is an ambiguity in
the choice of scale at which one renormalizes the SUSY masses versus the
scale at which one minimizes the potential, leading to uncertainties in the
one-loop contributions that can be large. The problem stems from the fact that
we minimize the potential at $Q=\mz$, while the SUSY masses are renormalized
at their thresholds. This leads to corrections of the one-loop potential that,
though of two-loop order, can become significant~\cite{lp3,klnpy}.
Nonetheless, because the solutions with large one-loop contributions are
disfavored by fine-tuning arguments, one can safely ignore this question.

The only significant deviation from Eq.~(\ref{ourfit:eq})
occurs in Fig.~\ref{mu:fig}b for
some points at very low $\mhalf$. These points correspond to the
$\mzero\gg\mhalf$ points in Fig.~\ref{envelopes:fig}
where the relic density was suppressed due to the presence of the $Z$- and
$\hh$-poles. These points can be missed in semi-analytic
approaches without calculating the neutralino relic density.
It is also this region which SUSY--$SU(5)$ proton lifetime
calculations like those of Ref.~\cite{an} select.
For these points we are far from the limit in which Eq.~(\ref{ourfit:eq}) holds
and one finds instead for $\mzero\gg\mhalf,\mz$\/ another tree-level
correlation between $\mu$ and $\mzero$~\cite{copw1} which fits our solutions
well (the effects of the one-loop effective potential here are very small):
\beq
\mu^2\simeq\mzero^2\,\frac{1+0.5\tan^2\beta}{\tan^2\beta-1}.
\label{copw:eq}
\eeq

Finally, one also sees from Fig.~\ref{mu:fig} that one can put
absolute upper and
lower bounds on $|\mu|$ within this framework, without regard to $M_2$. We
have included these $\mtpole$-dependent bounds in Table~\ref{bounds:table} and
Fig.~\ref{mu:fig}.

\subsection{Neutralino Relic Abundance and Dark Matter}

We have already emphasized the crucial role played by cosmological constraints
in deriving upper bounds of ${\cal O}(1\tev)$
on the supersymmetric mass parameters, {\em without} having to impose
a somewhat arbitrary constraint of fine-tuning. The main
ingredients that lead to such upper bounds in the CMSSM are: {\em (i)} the
lightest neutralino $\chi$ comes out to be the only possible LSP which
is neutral; and {\em (ii)} it turns out to be predominantly bino-like.

Even after rejecting solutions with charged LSPs as dark matter (DM)
candidates (in our case it is
$\widetilde\tau_1$ in the region $\mhalf\gg\mzero$),
the sneutrino could still have come out to be
the (neutral) LSP. However, just as in the more general case without Yukawa
coupling unification~\cite{kkrw1}, we never find that to be the case
after we apply experimental limits.
Furthermore, the neutralino LSP comes out to be almost pure bino
($p_{\bino}\gsim 95\%$ except for the region of $\mzero\gg\mhalf$ where
$p_{\bino}\gsim 90\%$) because of the radiative EWSB requirement
which effectively
leads to $|\mu|\gsim M_2$ (see Fig.~\ref{mu:fig}).

Finally, it should be appreciated (even if it has been known for some
time) that a neutralino relic abundance $\abundchi$ close to unity
corresponds to sfermion masses in the range of a few hundred \gev,
which is a natural mass scale in the MSSM with softly broken SUSY and
radiative EWSB. This fact makes the neutralino an excellent candidate for
the dark matter in the universe.
\begin{figure}
\centering
\epsfysize=2.75in
\hspace*{0in}
\epsffile{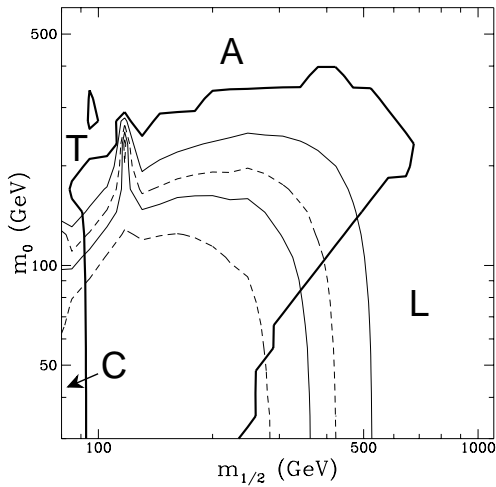}
\caption{The regions of the $(\mhalf,\mzero)$ plane consistent with
low $\tanb$ \btau\ mass unification, given all the constraints of the
CMSSM, for $\mtpole=170\gev$, $\azero/\mzero=0$ and $\mu<0$.
Solutions outside the thick solid lines are excluded: on the left (small
$\mhalf$) by the chargino mass bound ({\bf C}) $m_{\chi^\pm}>47\gev$ and
by tachyonic $\stp$'s ({\bf T}); on the right
(large $\mhalf\gg\mzero$) by charged LSP ({\bf L}); and from above by the age
of the universe, \ie\ $\abundchi\leq1$ ({\bf A}). We also indicate
the sub-regions selected by either the hypothesis of cold dark matter
($0.25\lsim\abundchi\lsim0.5$, between thin solid lines) or the one of
mixed dark matter ($0.16\lsim\abundchi\lsim0.33$, between thin dashed lines).
}
\label{dm:fig}
\end{figure}

Since in this approach all the masses and couplings are determined
in terms of just a few basic parameters, we can also reliably calculate
$\abundchi$ as a function of those same input parameters. We include all the
final states in calculating the neutralino pair annihilation even though the
dominant contribution in most of the parameter space (away from the poles)
comes from the exchange of the (lightest) sfermions.

We reject those solutions
for which $\abundchi>1$ as corresponding to the universe being too young
(less than about 10 billion years). This requirement alone appears to be
extremely powerful, excluding values of $\mhalf$ and $\mzero$ bigger than
roughly 1\tev\ and thus making {\em low-energy}
supersymmetry a {\em unique} outcome of the simplest SUSY grand-unification
assumptions. This is illustrated in Figs.~\ref{envelopes:fig}
and \ref{dm:fig}.

Furthermore, there is growing evidence for the existence of dark matter
in the universe. While its amount and nature remain unclear, one
of the most favored scenarios has been a flat universe ($\Omega=1$) with
most of its matter (about 95\%) contributed by DM. In one popular
scenario the neutralino would be the predominant component of such (cold) DM
in which case its relic abundance would be expected to be
\begin{equation}
\label{abundrangecdm:eq}
0.25\lsim\abundchi\lsim0.5.\ \ \ \ \ \ \ \ \ ({\rm CDM})
\end{equation}
More recently (after COBE),
a mixed CDM+HDM picture (MDM) has gained more attention as apparently
fitting the astrophysical data
better than the pure CDM model.
In the mixed scenario one assumes about
30\% HDM (like light neutrinos with $m_\nu\simeq 6\ev$) and about
65\% CDM (bino-like neutralino), with baryons contributing
the remaining 5\% of the DM.
In this case the favored range for $\abundchi$ is approximately given by
\begin{equation}
\label{abundrangemdm:eq}
0.16\lsim\abundchi\lsim0.33.\ \ \ \ \ \ \ \ \ ({\rm MDM})
\end{equation}
While neither of these DM scenarios is free from problems, it is nevertheless
interesting to point out which regions
of the parameter space of the CMSSM they select. This is
illustrated in Fig.~\ref{dm:fig} and in Table~\ref{DMbounds:table}.
We see in Fig.~\ref{dm:fig} that requiring either~(\ref{abundrangecdm:eq})
or (\ref{abundrangemdm:eq}) results in selecting only relatively narrow bands
in the ($\mhalf, \mzero$) plane. Their shape
and location vary with other parameters but typically correspond to
both $\mhalf$ and $\mzero$ in the range of a few hundred~\gev, independent
of the choice of $\azero$ and $\sgnmu$.
The resulting mass ranges are presented in Table~\ref{DMbounds:table} for
$\mtpole=170\gev$. They should be compared with those listed in
Table~\ref{bounds:table} for the same $\mtpole$ to appreciate how much more
limited the mass ranges become after the MDM/CDM assumption is made.
It is clear that, with the exception
of the light Higgs $\hh$, the mass spectra consistent with either CDM
or MDM are typically
beyond the current experimental reach. Conversely,
a discovery of a slepton at LEP~II, or a squark
(other than the stop) or gluino at the Tevatron well
below the limits given in Table~\ref{DMbounds:table},
while providing unquestionable evidence
for supersymmetry, would at the same time indicate clear deficiency
in the neutralino relic abundance~\cite{chiatlep2} below
the expected ranges of~(\ref{abundrangecdm:eq}) or~(\ref{abundrangemdm:eq}),
in the scenario with \btau\ unification and small $\tanb$.

\begin{table}
\centering
{\small
\begin{tabular}{|c||c|c||c|c||c|}
\hline
Mass Limits & \multicolumn{2}{c||} {MDM}  &
\multicolumn{2}{c||}{CDM} & MDM/CDM \\ \cline{2-5}
(GeV) & lower & upper & lower & upper & $+$~FT  \\
\hline\hline
$\mhalf$ & 90 & 520 & 90 & 660 & 230  \\ \hline
$\mzero$ & 55 & 245 & 85 & 245 & 245 \\ \hline
$|\mu(\mz)|$ & 270 & 940 & 280 & 1120 & 560 \\ \hline
$M_2$ & 80 & 430 & 80 & 545 & 190  \\ \hline
$\hh$ & 62 & 113 & 62 & 118 & 98 \\ \hline
$\hA$ & 370 & 1270 & 380 & 1530 & 770  \\ \hline
$\stp_1$ & 95 & 830 & 82 & 1020 & 430  \\ \hline
$\stau_1$ & 94 & 250 & 110 & 310 & 250  \\ \hline
$\widetilde{l_L}$ & 105 & 380 & 120 & 480 & 260  \\ \hline
$\widetilde q$ & 255 & 1110 & 265 & 1350 & 550  \\ \hline
$\gluino$ & 250 & 1200 & 250 & 1480 & 570 \\ \hline
$\chi=$\/LSP  & 25 & 225 & 25 & 290 & 100  \\ \hline
$\chi_1^\pm\simeq\chi_2^0$ & 48 & 435 & 48 & 550 & 200  \\ \hline\hline
$\tanb$ & 1.4 & 1.7 & 1.4 & 1.8 & 1.5  \\ \hline
$\alphas(\mz)$ & 0.123 & 0.128 & 0.123 & 0.128 & $0.125^\dagger$ \\ \hline
\multicolumn{6}{l}{\footnotesize ${}^\dagger$~Lower bound with FT constraint.}
\\
\end{tabular}}
\caption{The bounds on the mass parameters, $\tanb$, and $\alphas$ of the MSSM
under the extra constraints imposed by cold dark matter (CDM) and
mixed dark matter (MDM) scenarios, for $\mtpole=170$. The last column
(MDM/CDM + FT) gives
the upper bound when either the CDM or MDM scenario is assumed and the
requirement of low fine-tuning ($f\leq50$) is additionally imposed.
Uncertainties in the values are discussed in text and in caption of
\protect\ref{bounds:table}.
}
\label{DMbounds:table}
\end{table}

Finally, as we have noted previously,
the tiny region $\mzero\gg\mhalf\simeq m_Z$, corresponding to the
$Z$- and $\hh$-pole annihilation of the LSPs, is the region favored by the
non-observation of proton decay in $SU(5)$-type models with minimal
Higgs sectors~\cite{an}.

\section{Conclusions}

The predictability of the CMSSM becomes significantly enhanced by an
additional assumption of \btau\ unification at least in the region
of small $\tanb$ which we have studied here. Our main conclusions can
be summarized as follows:
\begin{enumerate}
\item
The parameter space of the CMSSM is now completely limited both from
below (by experimental constraints) and from above ($\mzero\lsim 500\gev$
and $\mhalf\lsim1\tev$) by robust cosmological constraints, without having
to invoke the a fine-tuning constraint.
This is a specific case of a more general property of the CMSSM: for
either small $\tanb$ (close to one) or for large $\mtpole\gsim170\gev$
the parameter space is {\em always} (\ie, for any choice of other parameters)
constrained from above broadly within the 1\tev\ mass range.
Both of these cases are selected by the requirement
of low $\tanb$ \btau\ unification.
\item
The resulting sparticle mass spectra are highly constrained and correlated.
All the colored sparticles (except for the lighter stop) are typically
very heavy, and so are the heavier Higgs bosons ($\hA$, $\hH$, $\hpm$).
The lightest neutralino is the LSP, and it is invariably predominantly
bino-like. Also, $\mcharone\simeq\mneuttwo\simeq2\mchi$.
The resulting mass ranges, with and without imposing the additional
fine-tuning constraint, have been listed in Table~\ref{bounds:table}.
It is clear that one can make a number of predictions which can falsify
the specific scenario considered here. It would be ruled out, for example,
if $\mtpole$ came out to be about 170\gev\ and the gluino or a squark (other
than a stop) were discovered at Fermilab well below 200\gev; similarly, the
sleptons cannot be much lighter than 65\gev.
\item
Additional stringent constraints are provided by requiring the neutralino
LSP to provide most of presumed dark matter in the flat universe, as we can
see from Table~\ref{DMbounds:table}. Again, we find
that the lower bound on the slepton masses (including the stau) are
beyond the reach of LEP~II, while the squarks and the gluino could
possibly
be discovered with the upgraded Tevatron in a limited region of
parameter space.
Thus, finding such sparticles
with masses well below the ranges given in Table~\ref{bounds:table}
would provide us with important information about the status
of the neutralino as the dominant component of DM in the universe.
\item
The predictions for \bsg\ in this scenario fall almost completely within the
range favored recently by CLEO, and near to the SM prediction. Furthermore,
for solutions with light spectra a sharp dependence
arises in the prediction of \bsg\ on $\sgnmu$. However, both theoretical and
experimental uncertainties must be reduced before one will be able to
constrain a large portion of the parameter space or determine $\sgnmu$
through this signal.
\item
In this restrictive scenario with imposed radiative electroweak symmetry
breaking an additional correlation between $\mu$ and $\mhalf$
arises (see Eqs.~(\ref{ourfit:eq})--(\ref{copw:eq})) which
may be helpful in a limited way in various phenomenological studies.

\end{enumerate}

Finally, which of the properties of the CMSSM sparticle spectra and
predictions are specific to the \btau\ unification assumption? Essentially,
the crucial ingredient is the requirement that $\tanb$ be close to one.
In this case the tree-level contribution to $\mhh$ is negligible and
$\hh$ is light enough to exclude large regions of the parameter space
for smaller $\mtpole$.
Also, the neutralino LSP pair-annihilation is genuinely somewhat suppressed
for $\tanb\simeq1$ leading to too much relic abundance ($\abundchi>1$) for
$\mhalf$ and $\mzero$ smaller than in the general case.
Allowing for larger $\tanb$ relaxes both lower
and upper limits on both $\mhalf$ and $\mzero$~\cite{kkrw1} and thus on the
sparticle and Higgs masses. Clearly, the sparticle mass spectroscopy of the
next generation of colliders will teach us a great deal about our theoretical
expectations, in particular on the question of \btau\ unification.

\section{Acknowledgements}
This work was supported in part by the U.S. Department of
Energy and the Texas National Research Laboratory Commission.




\begin{thebibliography}{99}
%
\bibitem{early}
P.~Langacker, in the {\em Proceedings of the PASCOS-90 Symposium}, eds.
P.~Nath and S.~Reucroft (World Scientific, Singapore, 1990);
P.~Langacker and M.-X.~Luo, \PRD{44}{91}{817};
J. Ellis, S. Kelley, and D.V.~Nanopoulos, \PLB{260}{91}{131};
U.~Amaldi, W.~de~Boer, and H.~F\"urstenau, \PLB{260}{91}{447};
F.~Anselmo, L.~Cifarelli, A.~Peterman, and A.~Zichichi, Nuovo Cim.
{\bf104A} (1991) 1817, and Nuovo Cim. {\bf105A} (1992) 581.
%
\bibitem{kkrw1}
G.L.~Kane, C.~Kolda, L.~Roszkowski, and J.~Wells, Michigan preprint
UM-TH-93-24 (October 1993), {Phys. Rev. \bf D, \it in press}.
%
\bibitem{bbop}
V.~Barger, M.S.~Berger, P.~Ohmann, and R.J.N.~Phillips, \PLB{314}{93}{351}.
%
\bibitem{cpw}
M.~Carena, S.~Pokorski, and C.~Wagner, \NPB{406}{93}{59}.
%
\bibitem{lp2}
P.~Langacker and N.~Polonsky, \PRD{49}{94}{1454}.
%
\bibitem{bothunif}
H.~Arason, D.~Casta\~no, B.~Keszthelyi, S.~Mikaelian, E.~Piard,
P.~Ramond, and B.~Wright, \PRL{67}{91}{2933}.
%
\bibitem{mbot}
H.~Arason, D.~J.~Casta\~no, B.~Keszthelyi, S.~Mikaelian, E.~J.~Piard,
P.~Ramond, and B.~D.~Wright, \PRD{46}{92}{3945};
S.~Titard and F.J.~Yndur\'ain, Michigan preprint UM-TH-93-25 (September 1993).
%
\bibitem{lepreview}
The LEP Collaborations ALEPH, DELPHI, L3, OPAL, and The LEP Electroweak
Working Group, CERN/PPE/93-157 (August 1993).
%
\bibitem{largetb}
B.~Ananthanarayan, G.~Lazarides, and Q.~Shafi, \PLB{300}{93}{245};
B.~Ananthanarayan and Q.~Shafi, Bartol preprint BA-93-25 (June 1993,
revised November 1993).
%
\bibitem{hrs} L.~Hall, R.~Rattazzi, and U.~Sarid, LBL preprint LBL-33997
(June 1993, revised February 1994).
%
\bibitem{copw:so10}
M.~Carena, M.~Olechowski, S.~Pokorski, and C.~Wagner, CERN preprint
CERN-TH.7163/94 (February 1994).
%
\bibitem{nelson}
A.~Nelson and L.~Randall, \PLB{316}{93}{516}.
%
\bibitem{lp3}
P.~Langacker and N.~Polonsky, Pennsylvania preprint UPR--0594T (February 1994).
%
\bibitem{an}
R.~Arnowitt and P.~Nath, \PLB{287}{92}{89}; \PRL{69}{92}{725};
\PLB{299}{93}{58}.
%
\bibitem{babubar}
K.S.~Babu and S.M.~Barr, \PRD{48}{93}{5354}; Bartol preprint BA-94-04 (1994).
%
\bibitem{abs}
B.~Ananthanarayan, K.S.~Babu, and Q.~Shafi, Bartol preprint BA-94-02
(January 1994).
%
\bibitem{flast}
H.~Baer, M.~Drees, C.~Kao, M.~Nojiri, and X.~Tata, Florida State preprint
FSU-HEP 940311 (March 1994).
%
\bibitem{hempfling}
R.~Hempfling and A. Hoang, DESY preprint DESY-93-162 (November 1993);
but see also J.~Kodaira, Y.~Yasui, and K.~Sasaki, Hiroshima preprint HUPD-9316
(November 1993).
%
\bibitem{nlc500higgsrep}
A.~Brignole, J.~Ellis, J.F.~Gunion,
M.~Guzzo, F.~Olness, G.~Ridolfi, L.~Roszkowski, and F. Zwirner, in the
{\it Proceeding of the Workshop ``$e^+e^-$ Collisions at 500\gev:
The Physics Potential''}, ed.~P.~Zerwas, DESY preprint 92-123A (August
1992).
%
\bibitem{higgsdetection}
Z.~Kunszt and F.~Zwirner, \NPB{385}{92}{3};
V.~Barger, M.S.~Berger, A.~Stange, and R.J.N.~Phillips, \PRD{45}{92}{4128};
J.F.~Gunion and L.H.~Orr, \PRD{46}{92}{2052};
H.~Baer, M.~Bisset, D.~Dicus, C.~Kao, X.~Tata, \PRD{47}{93}{1062};
H.~Baer, M.~Bisset, C.~Kao, X.~Tata, Florida State preprint
FSU-HEP-940204 (February 1994).
%
\bibitem{gunion}
J.~Gunion and H.~Pois, U.C.-Davis preprint UCD-94-1 (January 1994).
%
\bibitem{bsg}
S.~Bertolini, F.~Borzumati, A.~Masiero, and G.~Ridolfi, \NPB{353}{91}{591};
R.~Barbieri and G.~Giudice, \PLB{309}{93}{86}.
See also F.~Borzumati, DESY preprint DESY-93-090 (August 1993).
%
\bibitem{copw1}
M.~Carena, M.~Olechowski, S.~Pokorski, and C.~Wagner, CERN preprint
CERN-TH.7060/93 (October 1993).
%
\bibitem{klnpy}
S.~Kelley, J.L.~Lopez, D.V.~Nanopoulos, H.~Pois, and K.~Yuan, \NPB{398}{93}{3}.
%
\bibitem{chiatlep2}
L.~Roszkowski, \PLB{278}{92}{147}.
%
\end{thebibliography}
\end{document}